\documentstyle[preprint,aps]{revtex}
\begin{document}
\draft
\title{
Chaotic Synchronization of Symbolic Information\\
in the Discrete Nonlinear Schr\"{o}dinger Equation.}
\author{C. L. Pando L.}
\address{
IFUAP, Universidad Aut\'onoma de Puebla,\\
Apdo. Postal J-48. Puebla,Pue. 72570. M\'exico.\\
DEA, Universt\`a di Brescia,\\
Via Branze 38, 25123, Brescia, Italy.} 
\vspace{1cm}
\maketitle
\begin{abstract}

We have studied  the 
discrete nonlinear 
Schr\"{o}dinger 
equation (DNLSE)
with on-site  defects and
periodic boundary conditions.
When the array dynamics becomes chaotic,
the otherwise quasiperiodic
amplitude correlations between the oscillators
are destroyed via a symmetry breaking instability.
However, we show that
synchronization of symbolic information of suitable 
amplitude signals 
is possible
in this hamiltonian system.

\end{abstract}

\vspace{1cm}
\narrowtext

\noindent{\it 1. Introduction.}
The subject of our study is a subtle form of synchronization
that occurs in the discrete nonlinear Schr\"{o}dinger equation
(DNLSE).  
The DNLSE describes,
among other systems, 
localized modes in long protein systems
, arrays of nonlinear mechanical pendula and the
propagation of light in
nonlinear optical waveguide arrays based on Kerr media \cite{review}. 
These optical arrays were studied
intensively in the past several years, mostly in the context
of intrinsec localized modes also known as breathers,
both theoretically
\cite{aceves96,lederer} and experimentally
\cite{lederer}.
The DNLSE also
arises in recent models of Bose-Einstein condensates
trapped in periodic potentials generated by optical
standing-waves \cite{carretero}. 

In hamiltonian dynamical systems,
the usual notion of synchronization cannot be applied
, since volume must  be preserved in phase space \cite{rulkov,piki}.
Indeed, the trajectories of coupled systems have to
collapse to the synchronization manifold \cite{rulkov,piki}, which is only
possible in dissipative systems. 
However, a particular form of synchronization in hamiltonian
systems was found in article \cite{zanete}, where 
the orbits of two coupled systems cover almost
the same region of the
individual phase space but , typically,
at different times \cite{zanete}.
In the present work, we find that another form of
synchronization
is possible in a hamiltonian system such as the DNLSE.
We refer to synchronization of symbolic information (SSI).
This concept
has been applied
recently to describe the simultaneous production of common information
by two coupled chaotic 
oscillators as diverse as a logistic map and an electronic
circuit \cite{corron}.
Our article has four parts. The DNLSE and a
new set of 
DNLSE solutions 
are discussed in the second part. 
In part three,
we consider the 
DNLSE with an on-site defect, where chaotic 
synchronization of symbolic information occurs.
Finally, in part four, 
we give the conclusions. 

\noindent{\it 2. The Model and a Family of Stationary Solutions.}
We will consider the DNLSE within the framework of optical
waveguides.
Let us consider an array of one-dimensional coupled
waveguides. The neighboring waveguides are separated
from each other by the same distance $d$ and ,therefore,
the coupling constant between these is the same. We consider
the case with lossless Kerr media.
Within the framework of the coupled mode
theory, the evolution of
$E_{n}$, the electric field envelope in the nth waveguide,
is given by the following equation \cite{aceves96}:
$ i\frac{\partial E_{n}}{\partial z} +
\beta_{n} E_{n}+C(E_{n-1}+E_{n+1}) +
\gamma \mid E_{n}^{2} \mid E_{n}=0$.
This system is assumed to have periodic boundary conditions.
In this equation, $\beta_{n}$ is the linear propagation constant
of the nth waveguide,
$C$ is the linear coupling coefficient and $\gamma$  is the
nonlinear parameter. By introducing the dimensionless field
$Q_{n}=\sqrt{\gamma/2C} E_{n} \exp (-i(\beta+2C)z)$, this equation
transforms into the
discrete nonlinear 
Schr\"{o}dinger 
equation (DNLSE) given by :  
\begin{equation}
i\frac{\partial Q_{n}}{\partial \zeta} +
\delta_{n}Q_{n} +
(Q_{n-1}+Q_{n+1}-2Q_{n}) +
2 \mid Q_{n}^{2} \mid Q_{n}=0 ,
\end{equation}
where $\zeta=C z$ is the normalized propagation distance
and $\delta_{n}=\frac{(\beta_{n}-\beta)}{C}$.
The hamiltonian equations related to the DNLSE can be
obtained by transforming into action-angle variables
($I_{n},\theta_{n}$), where 
$Q_{n}=\sqrt{I_{n}}\exp(-i\theta_{n})$.
The quantity $I_{n}$ 
stands for the light intensity
of the nth waveguide \cite{review}. 

Now, we will consider
a family of 
stationary solutions for which
$\delta_{n}=0$.
To find these, we deal with the nonlinear
map approach \cite{hennig,bountis}. We underline that
there are different ways to find 
stationary solutions of the DNLSE \cite{review}.
In article \cite{hennig,bountis},
it was shown that 
homoclinic and heteroclinic orbits of a suitable 
hamiltonian map
support breather  solutions in the DNLSE.
Instead, 
we will consider those
stationary solutions which
are determined by the (elliptic)
stable periodic 
orbits of the hamiltonian  map mentioned above. 
These two types of DNLSE solutions are clearly
different. Indeed,
these  elliptic and homoclinic orbits , as it is well known,
have different origins and properties \cite{licht}.
While the first are periodic, according to the Poincar\'e-Birkhoff
theorem, the homoclinic orbits have no periodicity,
i.e., repeated iteration of the associated map produces new
homoclinic points \cite{licht}.
These  new stationary solutions
are obtained by making 
$\frac{d P_{n}}{d \zeta}=0$ and $\theta_{n}=\theta_{m}$
for any $n \neq m$. 
Moreover, we can define
the frequencies on these  tori $I_{n}=P_{n}^{2}$
by setting $\frac{d \theta_{n}}{d \zeta}=\lambda$,
where $\lambda$ is a  parameter.
Hence, the  following map 
is obtained : 
$X_{n+1}=P_{n}$ ;
$P_{n+1}=(\Gamma_{n}-2\mid P_{n} \mid^{2})P_{n}-X_{n}$,
where $\Gamma_{n}=2-\lambda-\delta_{n}$. 
The  Jacobian $J$
of this map is area preserving , i.e. $J=1$.
For the sake of definition, we will refer to this map 
as the cubic map (CM). 
To find our stationary solutions,
we will consider the case 
$\Gamma=2-\lambda$ for which $\delta_{n}=0$.
This map has the symmetry
$X_{n} \rightarrow -X_{n}$ and
$P_{n} \rightarrow -P_{n}$.
The fixed points of the CM are :
$(0,0)$ and $(\pm \sqrt{\Gamma/2-1}, \pm \sqrt{\Gamma/2-1})$.
When $\Gamma=2.5$,
we find a saddle point at $(0,0)$ and two elliptic points at
$(\pm \sqrt{\Gamma/2-1}, \pm \sqrt{\Gamma/2-1})$.
In Fig.~1a, 
quasiperiodic orbits
surround the elliptic point
$(\sqrt{\Gamma/2-1}, \sqrt{\Gamma/2-1})$ 
and further away,
we identify a period seven island chain and its resonances.
Surrounding these resonances, we also observe 
the characteristic chaotic sea in this nonintegrable
map \cite{licht}. 
We choose the stationary field $P_{n}$ 
associated with  the resonance of period seven. This 
is shown in Fig.~1b.
The stability of
$P_{n}$ was
numerically studied by integrating the DNLSE.
We have carried out these integrations 
with slightly different
initial conditions at $\zeta=0$.
That is, 
$\mid Q_{n}(\zeta=0)\mid = P_{n}+\nu_{n}$
, where $\nu_{n}$ stands for a small random perturbation
between
$-\nu_{max} < \nu_{n} < \nu_{max}$ and 
$\mid \theta_{n}(0) \mid \ll 1$ for any $n$.
In our calculations
$\nu_{max}= 10^{-3}$.
We find that the standard deviation of  
$I_{j}(\zeta)$ is proportional to the initial deviation
from the stationary solution.
This is a result of the stability of this stationary solution.
The finite number of
power spectrum components of $I_{j}(\zeta)$
indicates that these oscillations
are not chaotic. 
Indeed, we numerically found that
all the Liapunov exponents 
$\Lambda_{n}$
of this system converge
towards zero. 
The power spectrum $S(F)$
of $I_{3}(\zeta)$
is shown in the dashed line of Fig.~2a, where
$F$ is the renormalized frequency. 

\noindent{\it 3. Synchronization of Symbolic Information (SSI).}
To find the aforementioned correlation effects,
we add a small defect to
the linear propagation constant of the central
waveguide, which in Fig.~1b has
precisely the largest amplitude, i.e. 
$\delta_{3} \neq 0$ and
$\delta_{n}=0$
for $n \neq 3$.
We make use of the stationary
solution $P_{n}$ of period seven
as the initial 
condition for this perturbed DNLSE.
To estimate the extent of coherence between 
the intensities of 
a given pair of waveguides ($I_{m}$,$I_{n}$) 
we consider the  linear cross correlation 
function $R$ of this waveguide pair. 
Typically, for a small enough 
positive defect  $0 < \delta_{3} \ll 1$, the intensities
are not chaotic and the
symmetric waveguide pairs
, i.e. 
$(I_{1},I_{5})$, 
$(I_{2},I_{4})$ and
$(I_{6},I_{7})$, become strongly correlated for a suitable 
$\delta_{3} > 0$, i.e. $R \approx 1$ in the three waveguide pairs
\cite{pando}.

However, in an interval of the negative
defect $\delta_{3} < 0$ ,
it is possible to induce hamiltonian chaos in this system.
In contrast to the previous case,
the onset of chaos is
manifested when the
intensity pairs 
$(I_{1},I_{5})$, 
$(I_{2},I_{4})$ and
$(I_{6},I_{7})$ are no longer correlated. 
The exponential divergence of nearby trajectories
explains why two otherwise intensity 
correlated waveguides , such as 
$I_{1}$ and $I_{5}$, lose
their relative symmetry when chaos arises. 
This divergence is shown in  Fig.~2b, 
in which each of the aforementioned
intensity pairs executes initially stable small amplitude oscillations
before displaying erratic behavior. 
This behaviour is similar to that near the coupling resonance
in the three-dimensional billiards problem, where a particle
bounces back and forth between a smooth and a periodically rippled
wall \cite{licht}.

In the chaotic regime, in spite of the lack of symmetry 
of the aforementioned intensity pairs,
a very interesting form of synchronization
takes place between certain signals emerging from different
pairs of waveguides.
To this end, let us define the variables :
$J_{15}=I_{1}-I_{5}$,
$J_{24}=I_{2}-I_{4}$ and
$J_{76}=I_{7}-I_{6}$.
Our simulations show that 
$J_{15}$,
$J_{24}$ and
$J_{76}$ have the same sign during most of the time.
In fact, the signs of $J_{mn}$ are different during a
negligible interval. 
This is clearly appreciated in Fig.~2c.
We would like to know if the collective behaviour of 
the signals 
$J_{mn}$ suggests some kind of synchronization.
The answer is affirmative  from the point of view of
synchronization of symbolic information \cite{corron}.
According to this notion, two arbitrary oscillators
are perfectly synchronized ,in an information sense, if they
produce the same information, i.e., symbols generated by one system
map one-to-one to symbols emitted by the other system. Strictly
speaking, this form of synchronization requires that the common
information be emitted at precisely the same time. 
In article \cite{corron}, the 
chaotic signals of two coupled systems 
are compared using their symbolic dynamics. 
In our case, Fig.~2c suggests that the signals $J_{15}$,
$J_{24}$ and $J_{76}$ exhibit equivalent information
at the same rate. This ,however, does not contradict
the fact that
the usual notion of synchronization \cite{rulkov,piki}
cannot be applied to hamiltonian
systems ,such as ours.
Indeed, in our system, we compare the
chaotic signals $J_{nm}$ using just 
their symbolic dynamics. This is along the lines of the notion of
event synchronization \cite{peter}, where the relative timings
of events in the time series of different physiological signals are
compared.

The signals $I_{j}$, where $j=1,..,7$, are chaotic
as suggested by
the broadband power spectrum  $S(F)$ of $I_{3}(\zeta)$
shown in the solid line of Fig.~2a.
To generate a symbolic sequence out from 
$J_{15}$,
$J_{24}$ and
$J_{76}$,  we replace the value of $J_{ij}$ by 
$"-1"$ or $"1"$ if  
$J_{ij} < 0$ or
$J_{ij} > 0$ respectively. Let us call these signals $K_{ij}$.
In addition,
the symmetry of the system indicates that 
$<J_{ij}>=<K_{ij}>=0$, where $<>$ stands for the sample average.
We have calculated the autocorrelation function (ACF) $C(S)$
for each of these signals,
where 
$S$ is the space lag
\cite{licht}. The length of these signals 
is roughly $\zeta \sim 10^{5}$.
The ACF $C(S)$ of  
$K_{15}$,
$K_{24}$ and
$K_{76}$ coincide exceptionally well for all space lags $S$ 
considered. Indeed, all of them collapse to the solid line of
Fig.~2d, where $\delta_{3}=-0.0078$.
In this figure, the same result was obtained for
$\delta_{3}=-0.0158$. In both cases,
the associated linear
cross correlation function $R$ for any pair of signals $K_{mn}$ 
is given by $R \approx 1$.
In contrast, the ACF $C(S)$ of 
$J_{15}$,
$J_{24}$ and
$J_{76}$ agree only at a qualitative level. Moreover, the first
zero of these ACF 
$C(S)$
occurs at the same lag $S$, since
these series change sign almost simultaneously
as shown in Fig.~2c.

It is remarkable how the 
chaotic synchronization of symbolic information (SSI) of
$J_{15}$,
$J_{24}$  and
$J_{76}$ 
is manifested in the dynamics of
$\Delta \theta_{mn} = \theta_{m}-\theta_{n}$
when $\delta_{3} < 0$.
Indeed, as shown in Fig.~3a, all the $\Delta \theta_{mn}$
are bounded when SSI 
occurs , i.e. when $R \approx 1$
for any pair of series $K_{mn}$.
This locking of the phases $\theta_{m}$ and $\theta_{n}$
, $\mid \theta_{m}-\theta_{n} \mid < 2\pi$ when
$\mid \theta_{m}(0)-\theta_{n}(0) \mid < 2\pi$,
is also referred to as  
phase synchronization \cite{piko96}. The latter has been extensively
studied in the context of coupled self-sustained chaotic oscillators
\cite{piki,piko96}. Moreover, 
when $\delta_{3} < 0$ and the onset of phase slips of 
$\Delta \theta_{mn}$ occurs,
degradation of SSI 
($ 0 < R < 1$) takes place.
This is observed in Fig.~3b and Fig.~3c for which 
$\delta_{3}=-0.25$ and 
$\delta_{3}=-1.5$ respectively. 
When the dynamics of
$\Delta \theta_{mn}$ looks like a random walk, 
SSI is no longer a typical feature 
of the series $J_{mn}$.  
This is clearly appreciated in Fig.~4a, where the trajectory , at 
discrete intervals of $\zeta$, is projected on the plane
($J_{15}$,$J_{24}$).
If SSI took place, the
dots of this plane were only in the first and third quadrant of the
coordinate plane as shown by Fig.~4b and Fig.~4c.
To further characterize SSI , let us consider the spectrum
of Liapunov exponents $\Lambda_{n}$. 
This is shown in Fig.~4d.
The DNLSE has two constants
of motion, namely  the norm and the hamiltonian and 
,moreover, this flow is
autonomous. This, along with the symmetry of the Liapunov exponents
of hamiltonian systems
\cite{licht}, suggests that at least
four Liapunov exponents are equal to zero. 
For $\delta_{3}=-0.0078$,
there is a single positive Liapunov exponent whose
magnitude is much larger than that of the
other positive exponents. 
However, for
$\delta_{3}=-0.25$ and 
$\delta_{3}=-1.5$ five positive Liapunov
exponents have roughly
the same order of magnitude as shown in Fig.~4d. 
Finally, in the context of the aforementioned BEC models,
the phase locking of 
$\Delta \theta_{mn}$ 
is associated with the superfluid regime.
However, when 
$\Delta \theta_{mn}$ 
is unbounded, the system is said to behave as an insulator
\cite{carretero}

\noindent{\it 4. Conclusion.}
We have studied a subtle form of chaotic synchronization
that arises in a hamiltonian dynamical system with several
degrees of freedom.
This behaviour emerges in
a family of solutions
of the DNLSE, which
have as initial conditions
the neighborhood  
of certain stationary solutions. 
The latter, in turn, 
are given by 
the resonances of  a suitable hamiltonian map.
In a small negative interval of the defect 
($\delta_{3} < 0$ and $\mid \delta_{3} \mid \ll 1$),
suitable chaotic signals, which are generated by  different
pairs of oscillators, synchronize in the information
sense. Indeed, almost the same binary symbolic dynamics 
is emitted simultaneously by these signals.
When this happens, phase synchronization of 
all oscillators follows and ,roughly, just a single positive
Liapunov exponent is present. 
For arbitrary initial conditions or parameters, typically,
synchronization of symbolic information
does not take place. It is possible to extend this approach to
lattices with periodic defects and cuadratic nonlinearities. 
In view of the wide applicability of the DNLSE, we believe that
the aforementioned synchronization effect is generic in lattice models. 

\noindent {\bf acknowledgements}

This work was supported by CONACYT-M\'exico and the
ICTP-TRIL programme. 
I would like to thank Costantino de Angelis and Daniele Modotto
for their hospitality and useful suggestions. I also would like to thank
interesting discussions with Hilda Cerdeira, Arkady Pikovsky
and Stefano Ruffo.
\begin{figure}
\caption{
(a) Plot of the stationary field $P_{n}$
versus $P_{n+1}$ for 
$\Gamma=2.5$.
(b) Plot of the stationary field
amplitudes $P_{n}$
versus waveguide index $n$ for the
period seven resonance and $\Gamma=2.5$.}
\end{figure}
\begin{figure}
\caption{
(a) Plot of $\log_{10}S(F)$ versus 
$\log_{10}(F)$ for 
$I_{3}(\zeta)$.
$\nu_{max}=10^{-3}$ ,
$\delta_{3}=-0.0078$ 
(solid line) and
$\delta_{3}=0$ 
(dashed line). 
(b) Plot of the intensities $I_{1}$ (solid line)
and $I_{5}$ (dashed line) versus $\zeta$, where
$\delta_{3}=-0.0078$.
(c) Plot of 
$J_{15}$ (solid line)
$J_{24}$ (dashed line)
$J_{76}$ (dotted line)
versus $\zeta$ for 
$\delta_{3}=-0.0078$.
(d) Plot of the ACF
$C(S)$ of 
$K_{15}(\zeta)$ for
$\delta_{3}=-0.0078$ (solid line) and
$\delta_{3}=-0.0158$ (dashed line). The 
function
$C(S)$ of $K_{15}$, $K_{24}$ and $K_{76}$
coincide exceptionally well.}
\end{figure}
\begin{figure}
\caption{(a) Plot of 
$\Delta \theta_{15}$,
$\Delta \theta_{24}$, and
$\Delta \theta_{76}$ 
versus $\zeta$ for 
$\delta_{3}=-0.0078$.
Plot of 
$\Delta \theta_{15}$ (solid line),
$\Delta \theta_{24}$ (dashed line) , and
$\Delta \theta_{76}$ (dotted line) 
versus $\zeta$ for  (b)
$\delta_{3}=-0.25$ and (c)
$\delta_{3}=-1.5$.}
\end{figure}
\begin{figure}
\caption{
Plot of 
$J_{15}$ versus 
$J_{24}$ and $J_{76}$
for 
(a) $\delta_{3}=-1.5$, 
(b) $\delta_{3}=-0.0078$ and 
(c) $\delta_{3}=-0.0158$.
(d) Spectrum of Liapunov exponents $\Lambda_{n}$ for
$\delta_{3}=-0.0078$ (solid line),
$\delta_{3}=-0.25$ (dashed line), and
$\delta_{3}=-1.5$ (dotted line).}
\end{figure}
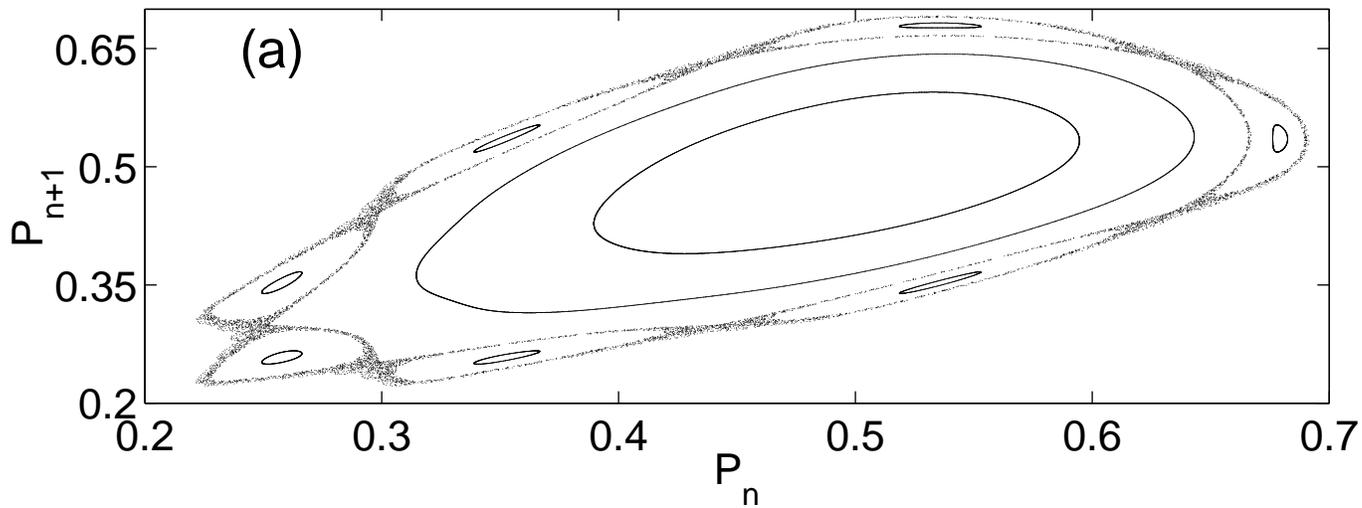
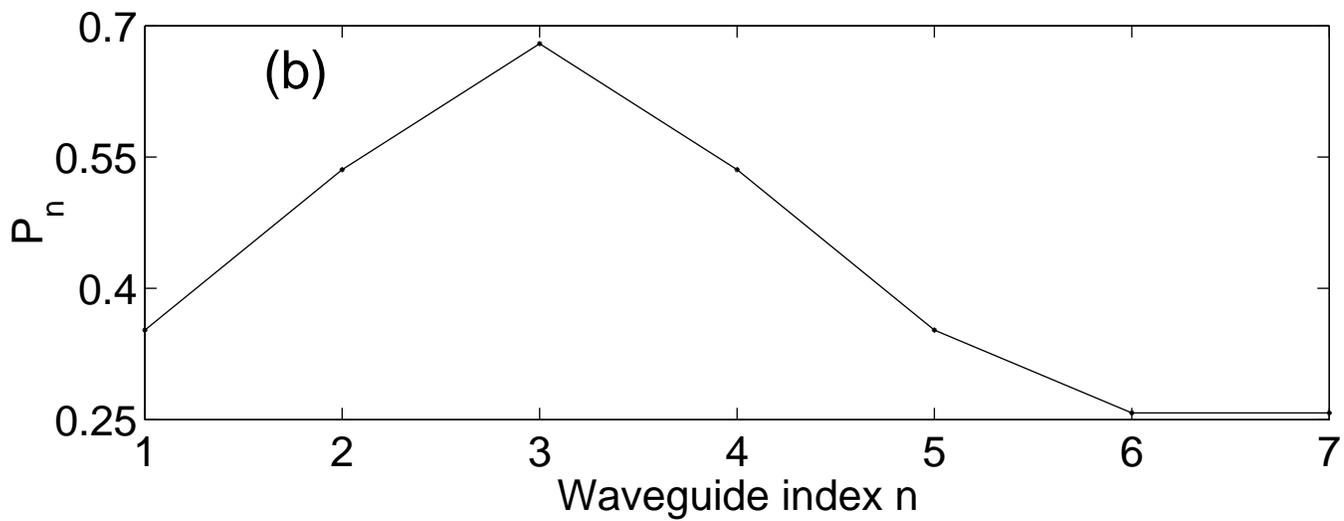
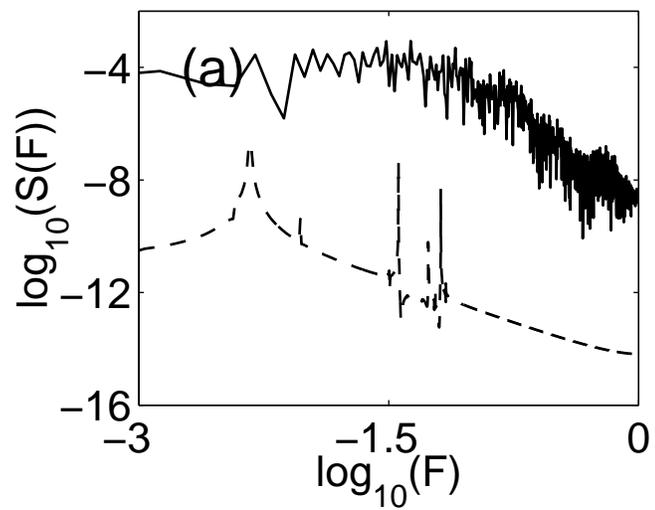
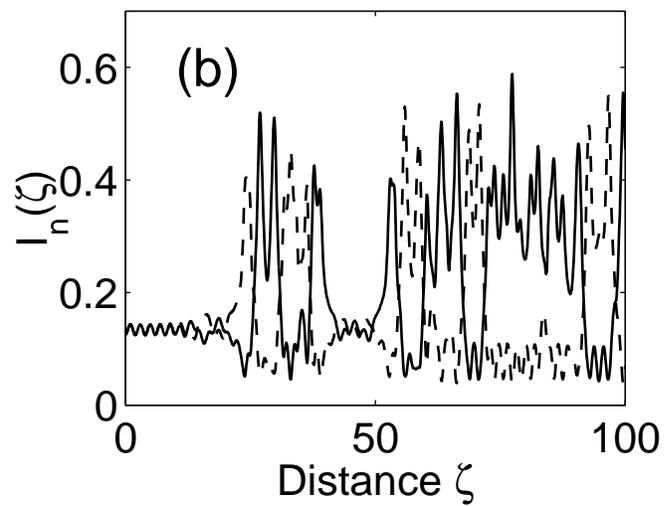
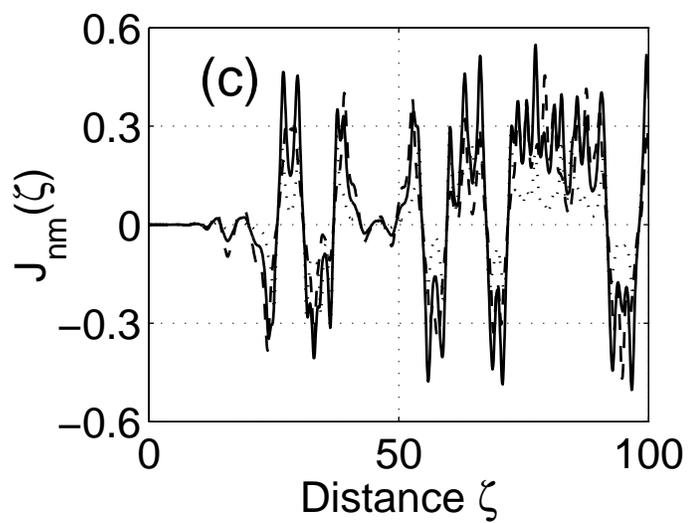
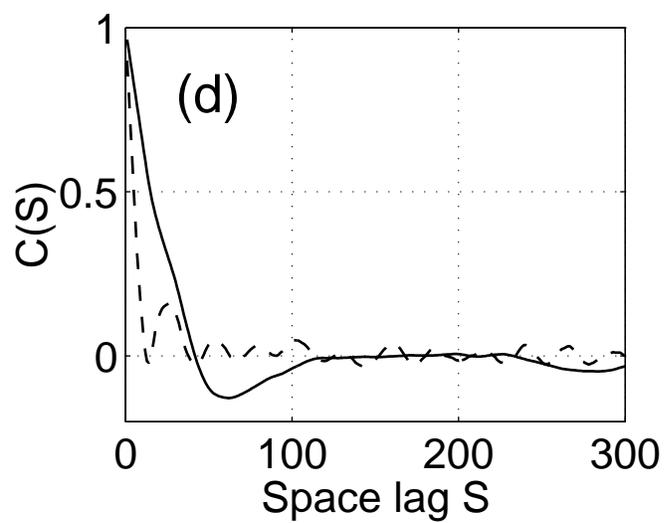
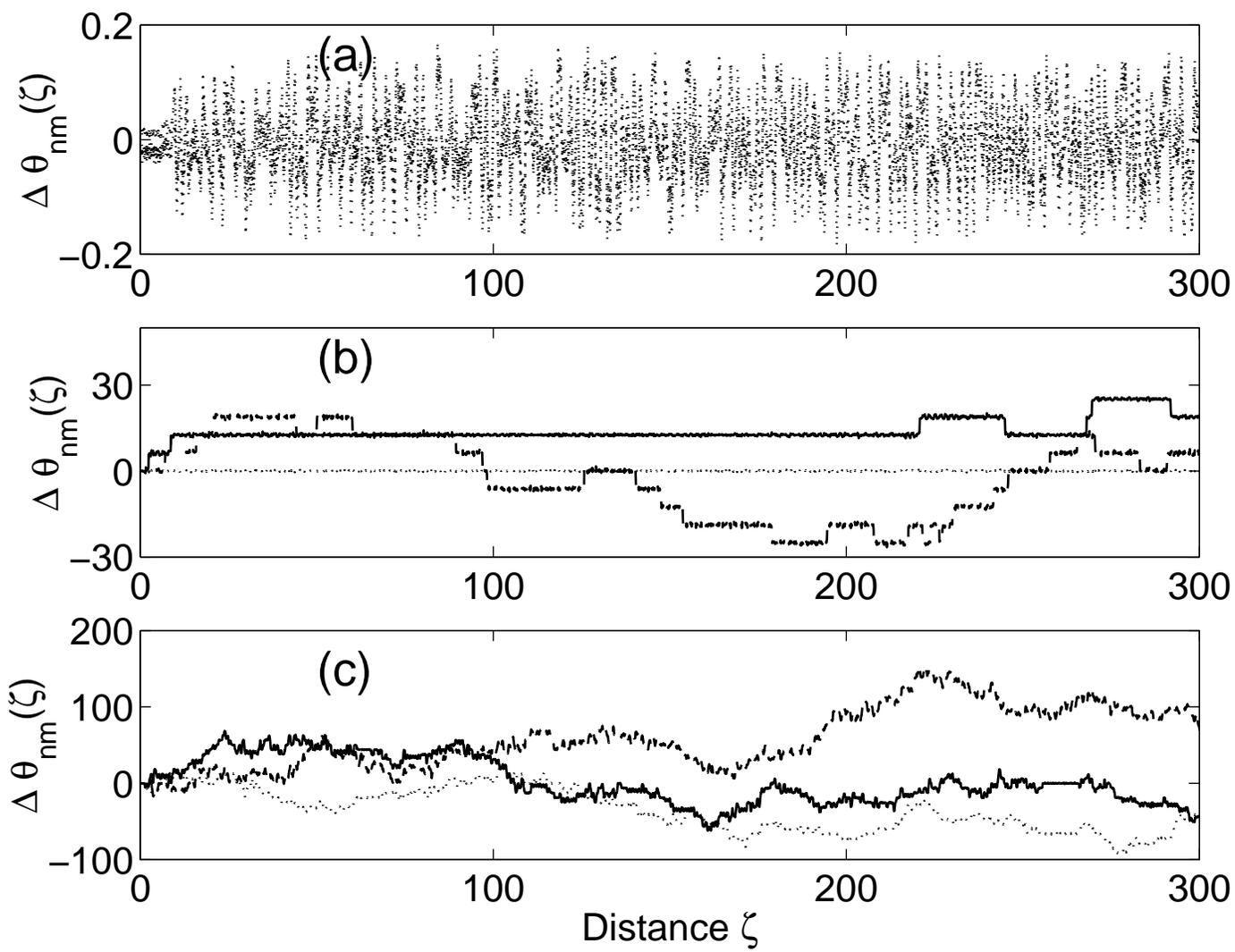
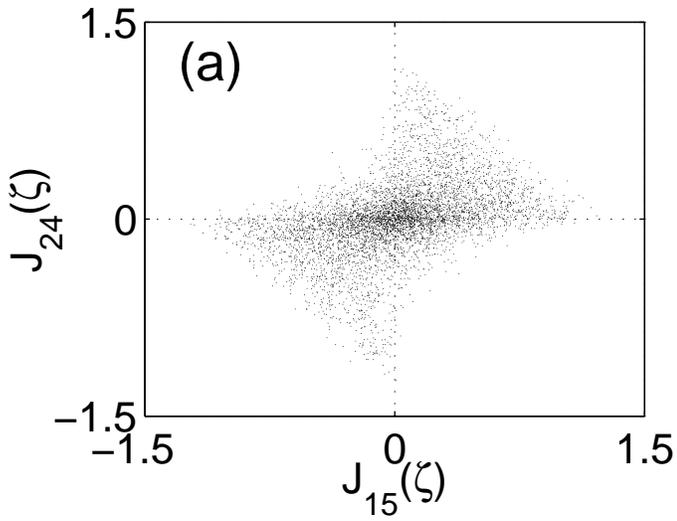
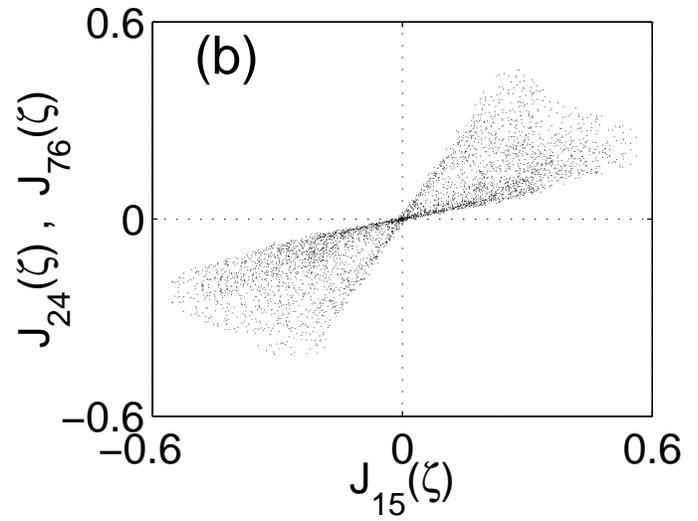
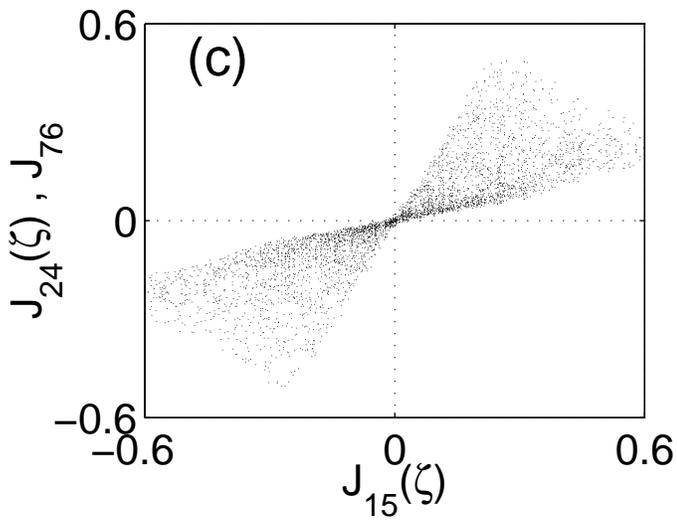
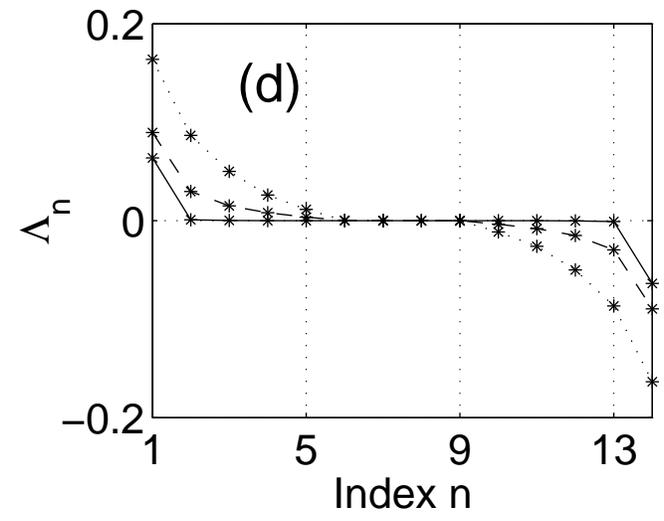
\end{document}